\documentclass[]{tMPH2e}
\usepackage{amssymb,amsmath}
\usepackage{graphicx}
\usepackage{epstopdf}
\newcommand{\angstrom}{\textup{\AA}}

\usepackage[english]{babel}
\begin{document}
\title{Constraining the size of the carrier of the $\lambda$5797.1 diffuse interstellar band}

\author{Jane Huang\footnote{Current affiliation: Harvard-Smithsonian Center for Astrophysics, 60 Garden St., Cambridge, MA 02138}\: and Takeshi Oka\footnote{Corresponding author. e-mail: t-oka@uchicago.edu} \\ Department of Chemistry and Department of Astronomy and Astrophysics, \\the Enrico Fermi Institute, University of Chicago, Chicago, Illinois, USA 60637}

\maketitle
\begin{abstract}
\label{abstract}

The diffuse interstellar band (DIB) at 5797.1 \AA\ is simulated based on three premises: (1) The carrier of the DIB is polar as concluded by T. Oka et al. from the anomalous spectrum toward Herschel 36. (2) The sharp central feature observed by P. J. Sarre's group is the $Q$-branch of a parallel band of a prolate top. (3) The radiative temperature of the environment is $T_r$~=~2.73~K.

 A $^2 \Pi \leftarrow$ $^2\Pi$ transition of a linear radical is simulated. Results depend on 10 parameters, with the rotational constant $B$ being the most critical. Comparisons of calculated spectra with observed data constrain $B$, which in turn constrains the number of heavy atoms to 5~$\leq n \leq 7$. Upper limits based on the Her 36 spectrum and lower limits based on stability against photodissociation are also discussed. The latter is based on the assumption that the DIB molecules are produced top-down from the breakdown of dust rather than bottom-up by chemical reactions.

A difficulty with this limit is that J. P. Maier's laboratory has observed many molecules within this size range, some of which have been tested astronomically. Candidates are discussed in light of many prolate tops observed by the P. Thaddeus and M. C. McCarthy microwave group.

\bigskip
\end{abstract}

\section{Introduction}

A long-standing question in astronomical spectroscopy
has been the identity of the carriers of the diffuse interstellar
bands (DIBs). DIBs are broad absorption features, primarily in the
visible region, that have been observed in the spectra of numerous
stars \cite{1995ARA&A..33...19H,IAU297}. After the incidental observation of the strongest DIB at 4430 \AA~ (henceforth referred
to as the $\lambda$4430 DIB) by Annie Jump Cannon in the early years of stellar spectroscopy \cite{Cannon,1958PASP...70..407C,1975ApJ...196..129H} and reports of the $\lambda\lambda$5780 and 5797 DIBs by Mary Heger in 1922 \cite{1922Heger,2013ProcRSocA...469...0604}, Paul W. Merrill demonstrated
in the 1930s that these spectral features arose due to intervening
interstellar material \cite{1934PASP...46..206M,1938ApJ....87....9M}.
Laboratory spectroscopy and computational modeling have led to a diverse
set of hypotheses for the origins of DIBs. Douglas argued
for the consideration of long-chain carbon molecules based on the recently discovered cyano-polyacetylene \cite{Avery} and the
Douglas effect \cite{Douglas} as a broadening mechanism \cite{1977Natur.269..130D}. On the other hand, van der Zwet and Allamandola proposed polycyclic aromatic hydrocarbons (PAHs),
based on their stability as well as their likelihood of forming free
radicals that could absorb in the visible region \cite{1985A&A...146...76V}. A similar proposal by L\'{e}ger and d'Hendecourt \cite{Leger} was published back to back with \cite{1985A&A...146...76V}.

Identifying specific DIB carriers would provide greater insight into
the nature of the interstellar medium (ISM). If the carriers were
known, the variations in DIB profiles along different lines of sight
could be used to infer physical conditions \cite{2006JMoSp.238....1S}.
Furthermore, DIBs have been detected in several moderate-redshift
galaxies, which has helped to constrain how long ago their carriers
first appeared and may provide information about the origins of organic
compounds in the universe \cite{2006ApJ...647L..29Y}.

Among the 400 or so DIBs observed thus far \cite{2008ApJ...680.1256H, 2009ApJ...705...32H}, the $\lambda\lambda$5780.5, 5797.1, 6196.0, 6379.3, and 6613.6 DIBs have been investigated in particular detail, both observationally and theoretically. The sharpness of these DIBs and substructures in some of these bands provide evidence that these DIBs are due to electronic transitions of gas-phase molecules \cite{1976MNRAS.174..571D,1995MNRAS.277L..41S,Walker}.
High-resolution spectra of these bands has facilitated rotational
contour fittings of DIB profiles, thereby enabling plausible sizes and geometries of carrier molecules to be assessed \cite{Walker,1996A&A...307L..25E,1996MNRAS.283L.105K,2000MNRAS.312..769S,2004ApJ...611.L113C}. In this paper we attempt to simulate the observed profile of $\lambda$5797.1, which has distinctive features that we believe are the most revealing of the properties of its carrier.

A remarkable recent development in  DIBs studies has been the discovery of the special sightline toward the star Herschel 36 by York's team in 2013, Dahlstrom et al. \cite{2013ApJ...773...41D}. Unlike the more than 200 stars previously observed, this sightline shows rotationally excited CH$^+$ and CH, revealing the high radiative temperature, $T$$_r$, of the environment. The effect of high $T$$_r$ is even more spectacular in DIBs; the $\lambda\lambda$5780.5, 5797.1, and 6613.6 DIBs show pronounced Extended Tails toward Red (ETR), while the $\lambda\lambda$5849.8, 6196.0, and 6379.3 DIBs are not much affected. They argued that the rotationally excited CH$^+$ and CH lines are due to far-infrared pumping by the nearby hot infrared source Her 36 SE discovered by Goto et al. \cite{2006ApJ...649..299G}, which is only 0.$''$25 away from Her 36. Oka et al. \cite{2013ApJ...773...42O} proposed that the ETRs are produced due to radiative pumping of the high rotational levels of polar carriers of the DIBs. This has introduced a new way to classify DIBs:  (1) those that have polar molecule carriers, which are sensitive to $T$$_r$, and (2) those with non-polar carriers, which are insensitive to variations in  $T$$_r$. Based on simulations of spectra of large linear molecules in which the rotational distribution was calculated by taking the radiative and collisional effects into account simultaneously, Oka et al. concluded that the carriers of the $\lambda\lambda$5780.5, 5797.1, and 6613.6 DIBs are polar molecules and that the ETRs are produced by a decrease in the rotational constant upon electronic excitation.

 In this paper, we use Oka et al.'s model for the rotational distribution of the carrier to simulate the $\lambda$5797.1 DIB. Earlier DIB modeling generally assumed a thermal Boltzmannian rotational distribution with a relatively high temperature, such
as \emph{T}~=~50~K by Danks and Lambert \cite{1976MNRAS.174..571D} and by Ehrenfreund and Foing \cite{1996A&A...307L..25E}, \emph{T}~=~8.9~-~101.3~K by Kerr et al. \cite{1996MNRAS.283L.105K}, \emph{T}~=~30~-~100~K by Schulz et al. \cite{2000MNRAS.312..769S}, \emph{T}~=~10~-~100~K by Walker et al. \cite{Walker}, and \emph{T}~=~21.0~-~25.5~K by Cami et al. \cite{2004ApJ...611.L113C}. This may give a reasonable approximation if the molecule is non-polar and the rotational distribution is governed by collisions. However, for polar molecules such as the predicted carrier type for the $\lambda$5797.1 DIB, radiative effects dominate. Spontaneous emission and low collision rates make the excitation temperature of the rotational distribution close to the cosmic microwave background blackbody temperature \emph{T}~=~2.73~K \cite{Lambert}. This reduces the width of a DIB and allows us to constrain the size of the carrier with higher accuracy. So far, Cordiner and Sarre's calculation of the $\lambda$8037~DIB with CH$_2$CN$^-$ as the carrier seems to be the only one in which $T$$_{ex}$~=~$T$$_r$~=~2.7~K is used \cite{Sarre}. Cossart-Magos and Leach \cite{1990A&A...233..559C} and Edwards and Leach \cite{1993A&A...272..533E} used thermal distributions with $T$$_{ex}$~=~3~K, although their studies were on non-polar molecules.

\section{Methods}

\subsection{The $\lambda$5797.1 DIB}

\begin{figure*}
\begin{center}
\includegraphics[scale=0.5]{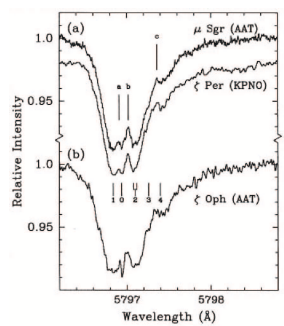}
\caption{Spectra (resolving power of 600,000 at AAT and 300,000 at KPNO) of
the $\lambda 5797.1$ DIB, originally published as Figure 2 in Kerr
et al. \cite{1998ApJ...495..941K}. Reprinted with permission of the main author (P. J. Sarre) and The Astrophysical Journal.}
\label{fig1}
\end{center}
\end{figure*}

The substructures of the $\lambda$5797.1 DIB were
first studied at very high resolutions by Sarre's group in the spectra of the stars $\mu$ Sgr, $\zeta$ Per, and $\zeta$ Oph \cite{1995MNRAS.277L..41S,1998ApJ...495..941K}.
Figure \ref{fig1} shows Kerr et al.'s spectra of the $\lambda 5797.1$ band
for three lines of sight, observed either at the Anglo-Australian Telescope (AAT)
at a resolving power of 600,000 or at Kitt Peak National Observatory
at a resolving power of 300,000 \cite{1998ApJ...495..941K}. As
seen in all three lines of sight, the $\lambda$5797.1 DIB shows a
narrow spectral feature at its center, with
two peaks on either side that are wider but at a similar
depth. On the redward side of these three peaks, there is a shallow
bump in the spectrum \cite{1998ApJ...495..941K}. Later high-resolution
studies of the $\lambda$5797.1 DIB along several
other lines of sight also consistently identified similar spectral
features \cite{2008MNRAS.386.2003G,2008ApJ...682.1076G}.

The characteristic three features of the $\lambda$5797.1 DIB, unique among known DIBs, are most naturally ascribed to the $R$-,  $Q$-, and $P$-branches of a fairly simple molecule. The sharp and weak central feature suggests a $Q$-branch of a parallel transition, since a perpendicular transition would feature a stronger and broader $Q$-branch reminiscent of the central peak of the $\lambda$6613.6 DIB. The carrier must be a prolate top, because otherwise the $Q$-branch would be stronger (see, for example, Cossart-Magos and Leach's rotational contour simulations of PAHs as carriers \cite{1990A&A...233..559C}). For the $Q$-branch to appear, the carrier of the $\lambda$5797.1 DIB should be either a linear molecule with non-zero orbital angular momentum $\Lambda$, or an asymmetric or symmetric rotor with a large $A$ rotational constant, such as carbene with C$_{2v}$ symmetry or methyl polyynes with C$_{3v}$ symmetry. In this paper, we simulate the $\lambda$5797.1 DIB as a $^2 \Pi \leftarrow$ $^2\Pi$ (parallel) transition of a linear molecule. Linear molecules with $\Delta$ ground states (i.e., $\Lambda$~=~2) or with higher $\Lambda$ are rare. The C$_{2v}$ and C$_{3v}$ molecules are not modeled in this paper, but the constraints on the rotational constant $B$ and hence the size of the molecule should apply similarly. During the process of simulation, we found that modeling linear molecules was advantageous in that an additional parameter is made available by including a large spin-orbit interaction that changes upon electronic excitation.

Oka et al. \cite{2013ApJ...773...42O} concluded that the ETRs for the $\lambda\lambda$5780.5, 5797.1, and 6613.6~DIBs observed toward Herschel 36 are caused by bond lengthening and hence a decrease of the rotational constant $B$ upon excitation of the electronic state. For such a transition, vibronic bands involving stretching vibrations may have large Franck-Condon factors. Therefore, the $\lambda$5797.1 DIB is likely accompanied by fairly strong vibronic progressions. About 60 DIBs have been reported within 1000 cm$^{-1}$ to the blue of the $\lambda$5797.1 DIB toward both HD 204827 \cite{2008ApJ...680.1256H} and HD 183143 \cite{2009ApJ...705...32H}. Identifying which of these many candidates are vibronic satellites of the $\lambda$5797.1 DIBl requires careful studies of correlations \cite{2011ApJ...727...33F}, progressions \cite{2010ApJ...712L.165D}, and spectroscopy. Further work in this area will be attempted as a separate project in future publications.

\subsection{Calculating rotational distributions}

The population distribution of rotational levels for a polar linear
molecule in this work was calculated using Oka et al.'s approach for
modeling the Her 36 DIBs \cite{2013ApJ...773...42O}. Accounting for
both radiation and collisions, the relationship between the populations
of the rotational levels in the $J\rightarrow J-1$ and $J\leftarrow J-1$
transitions is determined by the principle of detailed balancing \cite{2013ApJ...773...42O}

\begin{equation}n(J)(A^J+B_{J-1}^\downarrow\rho+C_{J-1}^\downarrow) = n(J-1)(B_{J-1}^\uparrow\rho+C_{J-1}^\uparrow),\end{equation}
\noindent where $A^J$ is the Einstein coefficient for spontaneous
emission, $B_{J-1}^\downarrow$ and $B_{J-1}^\uparrow$ are the
Einstein coefficients for stimulated emission $J~\rightarrow~J-1$ and absorption $J~\leftarrow~J-1$, respectively, $\rho$
is Planck's spectral energy density, and $C_{J-1}^\downarrow$ and $C_{J-1}^\uparrow$
are the collision rates.

From Equation (1), the following expression for number density was obtained by Oka et al. \cite{2013ApJ...773...42O}\cite{erratum}

\begin{equation}
n(J) = n(0)\prod_{m = 1}^{J} \frac{\alpha B^3 \mu^2\frac{m^4}{2m-1}\frac{1}{\exp(2hBm/k T_r)-1} + C\sqrt{\frac{2m+1}{2m-1}}\exp\left(\frac{-hBm}{kT_k}\right)}{\alpha B^3 \mu^2\frac{m^4}{2m+1}(1+ \frac{1}{\exp(2hBm/k T_r)-1}) + C\sqrt{\frac{2m-1}{2m+1}}\exp\left(\frac{hBm}{kT_k}\right)},\end{equation}
\noindent where $\alpha$~=~2$^9\pi^4/3hc^3$, $\mu$ is the permanent dipole moment
of the molecule, $B$ is the molecule's rotational constant, $T_r$ is the radiative temperature,
$T_k$ is the kinetic temperature, and $C$ is the averaged collision rate.

Values of $n$($J$) calculated from Equation~2 depend critically on $T_r$ and $B$, but not as much on $T_k$, $C$, and $\mu$ within the range of typical values. Thus, as in Oka et al.,  we set $C$~=~10$^{-7} ~$s$^{-1}$, $\mu$~=~4 Debye, and $T_k$ = 100 K \cite{2013ApJ...773...42O}; the collision rate is based on a number density of $n~\sim$~100~cm$^{-3}$ and collision rate constant $\langle\sigma v\rangle~ \sim~ 10^{-9}$ cm$^3$ s$^{-1}$, while the kinetic temperature is chosen based on the typical rotational temperatures of 45~K to 200~K for H$_{2}$ in interstellar clouds \cite{1986ApJS...62..109V}. It is readily shown that $n$($J$) reduces to the Boltzmann distribution with radiative temperature $T_r$ if collisional effects are negligible ($C$~=~0), and to that with kinetic temperature $T_k$ if the molecule is non-polar ($\mu$~=~0).

The assumed permanent dipole moment of 4 Debye is typical for molecules considered in this paper; see for example Maluendes and McLean \cite{McLean} for H$_2$C$_n$ and Woon \cite{Woon} for HC$_n$. Varying the dipole moment by factors of a few does not affect the calculation much because of the $J^3$ factor in the Einstein coefficient $A^J$.

\subsection{Calculating spectral profiles}

We use a simplified formula for the energy of spin-orbit interaction and rotation for  a $^2\Pi$ electronic state, assuming Hund's case (a) \cite{Landau}

\begin{equation} E~=~A\Sigma\Lambda~+~BJ(J+1), \end{equation}

\noindent where $A$ is the spin-orbit splitting constant and $\Sigma$ ($\pm$~1/2) and $\Lambda$ ($\pm$~1) are the projections of spin angular momentum $\bf{S}$ and orbital angular momentum $\bf{L}$, respectively, onto the molecular axis. The spin-orbit interaction splits the $^2\Pi$ state into two states with total electron angular momentum  $\Omega$~=~$\Sigma$~+~$\Lambda$ = ($\pm$~3/2) and ($\pm$~1/2). These two states are designated as $^2\Pi_{3/2}$ and $^2\Pi_{1/2}$, with energy $A$/2 and $-A$/2, respectively.
Smaller effects such as centrifugal distortion ($D$), $\Lambda$ doubling ($q$), spin-rotation interaction ($\gamma$), and hyperfine interaction ($a$) are neglected. From the energy formula, the line positions are given by:
\begin{subequations}\label{grp1}
\begin{align}
P(J): &  ~\nu = \nu_0\pm\frac{1}{2}(A'-A)+B'J(J-1)-BJ(J+1)\label{first}\\
Q(J): & ~\nu = \nu_0\pm\frac{1}{2}(A'-A)+(B'-B)J(J+1)\label{second}\\
R(J): & ~\nu = \nu_0\pm\frac{1}{2}(A'-A)+B'(J+1)(J+2)-BJ(J+1)\label{third},
\end{align}
\end{subequations}
\noindent where $\nu_0$ is the band origin and +($A'-A$) and $-$($A'-A$) apply to the $^2\Pi_{3/2}$ and $^2\Pi_{1/2}$ states, respectively.
We denote quantum numbers and molecular constants in the electronic excited states with primes to distinguish from the ground state, which is unprimed.

The relative intensity of a line in an electronic band is given by
\cite{Herzberg}

\begin{equation}I \propto \frac{n(J)S}{2J+1}, \end{equation}

\noindent where $S$ is the appropriate H\textup{\"o}nl-London factor, or line strength. For Hund's case (a), the H\textup{\"o}nl-London factors are given by \cite{Kovacs}

\begin{subequations}\label{grp2}
\begin{align}
P(J): &  ~\frac{J^2-\Omega^2}{J}\label{first}\\
Q(J): & ~\frac{\Omega^2(2J+1)}{J(J+1)}\label{second}\\
R(J): & ~\frac{(J+1)^2-\Omega^2}{J+1}\label{third}.
\end{align}
\end{subequations}

Equation~6b indicates that the $Q$ branch line of $^2\Pi_{3/2}$ is 9 times stronger than that of $^2\Pi_{1/2}$. Therefore, the strong $Q$ branches in Figure~1 indicate that $^2\Pi_{3/2}$ and $^2\Pi_{1/2}$ are irregular, that is, $A$ is negative \cite{Herzberg}.

Lines were computed for values
of $J$ up to 99, with the assumption that populations in higher levels
could be neglected. As $J$ increases, Hund's case (a) gradually shifts to Hund's case (b). However, for the large molecules considered in this paper's simulations, the value of $J$ at which this occurs is so large that
the use of Hund's case (a) for the energy and intensity formulae does not introduce serious errors.

Each spectrum is simulated by using the molecular population given in Equation~2, frequencies calculated from Equations~4a-4c, and the corresponding intensity formulae based on Equations~5 and 6a-6c. Individual lines are broadened by the uncertainty due to the limited time of spontaneous emission and internal conversion \cite{Douglas}. This results in a Lorentzian lineshape with linewidth $\Gamma=\frac{1}{2\pi\Delta t}$, where $\Delta t$ is the lifetime of the level. Instrumental broadening of the lines was approximated by a Gaussian with a standard deviation of $\sigma = \frac{\nu_0}{2\sqrt{2\ln 2}R}$,
where $\nu_0$ is the band origin in Table 1 and $R$ is the instrument resolution. The combination of the effects of lifetime and instrumental
broadening results in a Voigt lineshape function.

\subsection{Simulating spectra}

There are nine variables in the above formalism: the spectroscopic constants $A'-A$, $B'$, $B$, the permanent dipole moment in the ground state $\mu$, the
radiative and kinetic temperatures $T_r$ and $T_k$, the average collision rate $C$, and the line parameters $\Gamma$ and $\sigma$. We use one more variable $\rho$, the ratio of the peak intensities of the $^2\Pi_{3/2}$~$\leftarrow$~$^2\Pi_{3/2}$ and the $^2\Pi_{1/2}$~$\leftarrow$~$^2\Pi_{1/2}$ components, which depends on the populations in each ground state. In principle, this can be calculated from $T_r$ and $T_k$. However, this would require analyzing the radiative effects of the magnetic dipole moment and magnetic collisions, which are not well known. We therefore leave $\rho$ as a free parameter. $\sigma$ is held fixed during simulations, since it depends only on the resolving power of the instrument. As noted in the previous section, $n(J)$ is relatively insensitive to changes in $T_K$, $\mu$, and $C$; these parameters were also held fixed. Because the spectrum of $\lambda$5797.1 toward the star 20 Aql was used as a reference during the simulations, $T_r$ was set to 2.73 K initially. The excitation temperature of CN toward 20 Aql has been measured to be 2.728 K \cite{Lambert}, and it is assumed that $T_{ex}$ = $T_r$. Thus, the quantities that were estimated through the modeling process were $B$, the ratio $B'/B$, $\Delta t$, and the difference in the spin-orbit coupling constants $A'-A$.

 Although the spectrum toward $\zeta$~Oph in Figure \ref{fig1} shows the $Q$-branch most clearly with the high resolution ($R$~$\sim$~600,000) of the AAT, the spectrum is not numerically available. Therefore we use the spectrum toward star 20 Aql reduced by D. E. Welty from data retrieved from the ESO Science Archive Facility (programme IDs 078.C-0403 for the 20 Aql observations by Galazutdinov et al. \cite{2008MNRAS.386.2003G}) as the reference spectrum to compare with our simulation. The High Accuracy Radial Velocity Planet Searcher (HARPS) spectrograph used has lower resolution ($R$~=~115,000) and the $Q$-branch is not as clearly observed, but the spectrum allows for comparisons of the spectral widths between the observed and simulated spectra.

All simulated DIBs were rescaled in the figures to have the same maximum absorption as the given reference spectrum.  The absence of Doppler
splitting of the interstellar atomic lines toward 20 Aql suggests
that the DIB carriers may reside in a single interstellar cloud, which would
allow for intrinsic profiles of the $\lambda$5797.1 band to be observed \cite{2008MNRAS.386.2003G}. Furthermore, since the DIBs toward 20 Aql are among the narrower ones found in published high resolution spectra  \cite{2013ApJ...773...41D}, using 20 Aql for reference facilitates making estimates of a lower bound for $B$.

\section{Results}

\subsection{Modeling the $\lambda$5797.1  DIB at high resolution}

\subsubsection{Variation of parameters}

\begin{figure*}
\begin{center}
\includegraphics{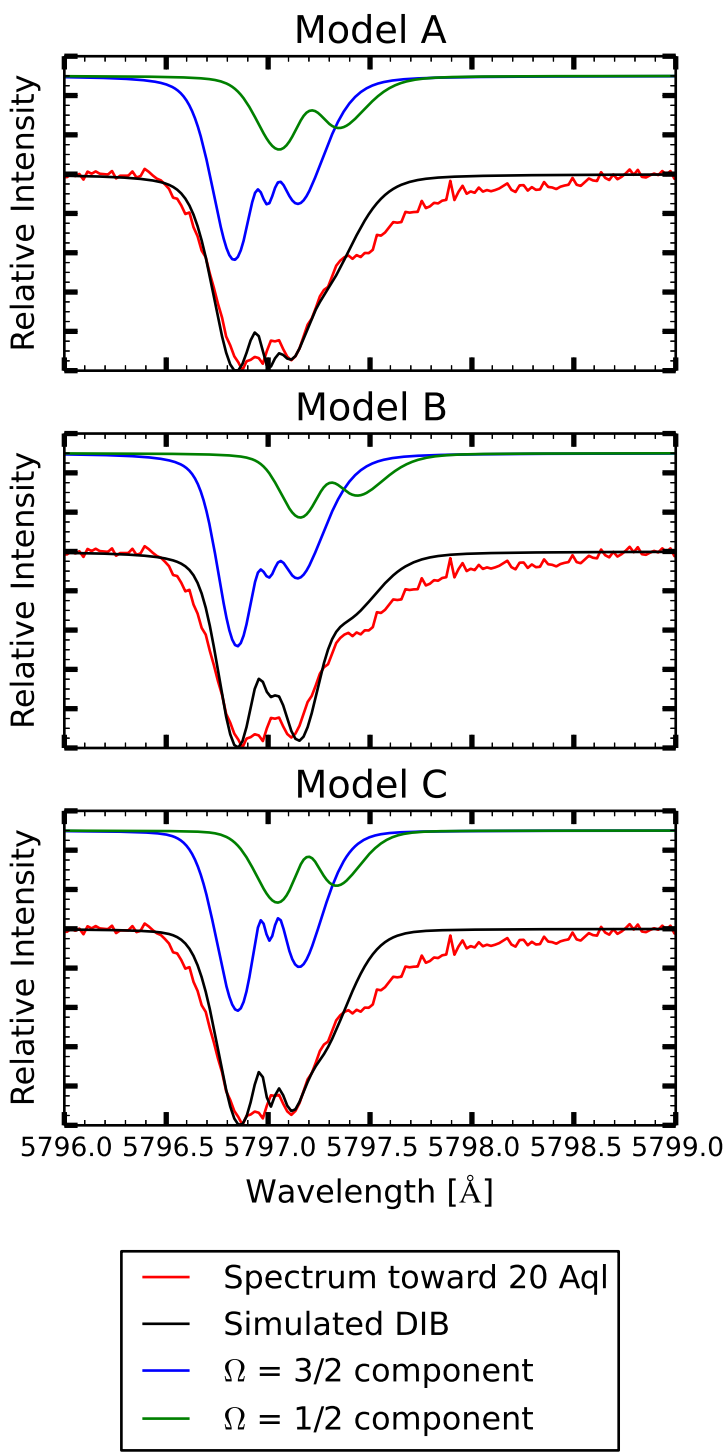}
\caption{Three examples of simulated spectra for the $\lambda$5797.1 DIB using a $^2 \Pi \leftarrow$ $^2\Pi$
transition. The observed 20 Aql spectrum (courtesy of D. Welty) is at $R=115,000$ and based on data obtained from the ESO Science Archive Facility. Models A, B, and C have been calculated using the three sets of parameters listed in Table \ref{table1}}
\label{fig2}
\end{center}
\end{figure*}

\begin{table*}[t]
\renewcommand{\arraystretch}{1.5}

\caption{Model input values for the $\lambda$5797.1 DIB (see fig. \ref{fig2})}

\label{table1}
\noindent \centering{}%
\begin{tabular}{c@{\hspace{5pt}}l@{\hspace{5pt}}l@{\hspace{5pt}}l@{\hspace{5pt}}l}
\hline
Model & A & B  & C\\
\hline
\hline\\[-2ex]
$T_r$ {[}K{]} & 2.73 & 2.73 & 3.50\\
$B$ {[}MHz{]} & 1300 & 1200 & 1000\\
$B'/B$ & 0.98 & 0.97 & 0.99\\
$A'-A$ {[}MHz{]} & 19000 & 27000 & 17000\\
$\lambda_0$ {[}$\angstrom${]} & 5797.10 & 5797.15 & 5797.10\\
$\Delta$t {[}ps{]} & 50 & 50 & 100\\
$\rho$ & 2.50 & 3.00 & 2.50\\
\hline

\end{tabular}
\end{table*}

The observed $\lambda$5797.1 DIB toward 20 Aql was simulated by varying parameters. Three examples are shown in Figures 2a, 2b, and 2c, referred to as Models A, B, and C, respectively. The corresponding parameters are given in Table \ref{table1}. Individual spin-orbit components are drawn above the simulated profile. Out of the 7 parameters shown in Table \ref{table1}, we found that the 4 parameters $T_r$, $B$, $A'-A$ and $\rho$ are crucial in reproducing the spectrum. Though this is too many to fix from a spectrum, we found that they are not tightly correlated. Thus, we could narrow down their values by varying each parameter individually. Model A gives a reasonable agreement, although the shallow bump on the redward side of the DIB is not reproduced. A simulation with larger $A'-A$, as shown in Model B, would shift the $\Omega=1/2$ spin-orbit component so as to create a small bump on the redward side; however, the central peak then becomes substantially weaker.  Increasing $T_r$ could compensate for the narrowing of profiles with smaller values of $B$, as shown in Model C with $T_r$ = 3.5 K. However, as noted earlier, $T_r$ should be close to 2.73~K, the CN excitation temperature for this sightline \cite{Lambert}. Thus, model A is more plausible.

\subsubsection{Variation of $B$}

\begin{figure*}
\begin{center}
\includegraphics[scale=0.6]{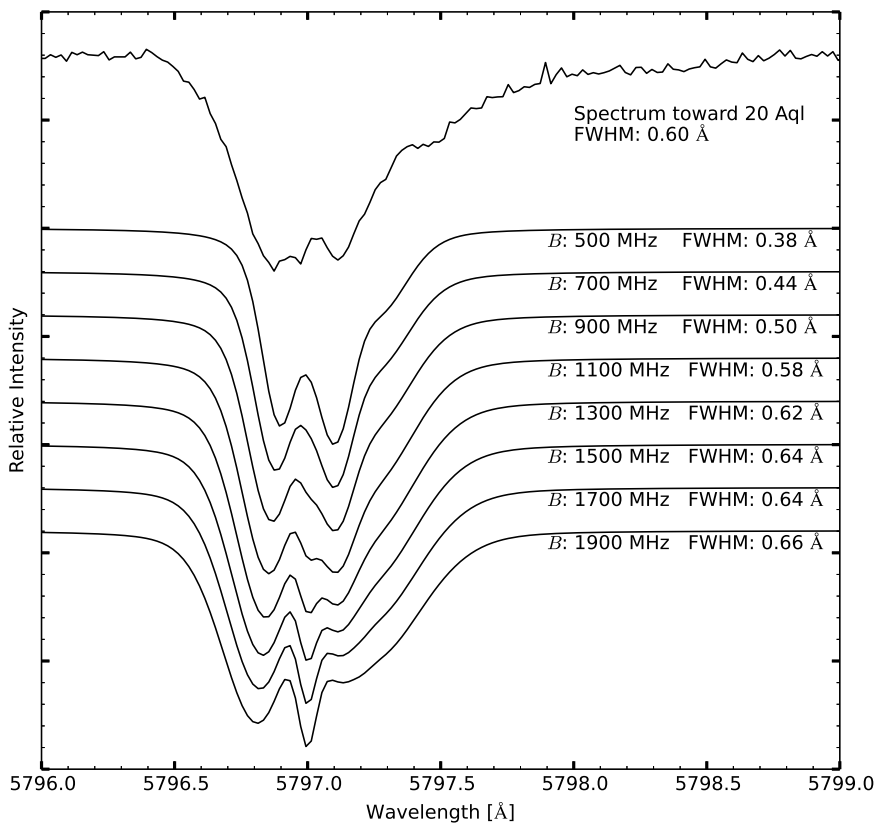}
\end{center}
\caption{Variation of the simulated spectra as a function of the rotational constant $B$. The agreement between the simulated and observed full width at half maximum (FWHM) and the visibility of the $Q$-branch sets the limit of 900 MHz $\leq$ $B$ $\leq$ 1900 MHz which approximately corresponds to 5 $\leq$ $n$ $\leq$ 7.}
\label{fig3}
\end{figure*}

The rotational constant $B$ is the key parameter in simulating the DIB and constraining the size of the carrier molecule. As noted in Section 3.2, $B$ and $T_r$ are two crucial parameters for determining the rotational distribution, but $T_r$ is fixed at 2.73~K. Figure \ref{fig3} shows the effect of changing $B$ on the profile of the $\lambda$5797.1 DIB with other parameters fixed at those in Model A. These results show that $B$ is most likely 900~MHz~\textless~$B$~\textless1900 MHz. For a lower $B$, the full width at half maximum (FWHM) is too small. In addition, the relative strength of the characteristic $Q$-branch would be too small to be observable because of the high $J$ values required by a smaller $B$. The line width can be made arbitrarily large by assuming a very small $\Delta t$, but then the sharp $Q$-branch would be washed out. For a larger $B$, the FWHM is too large and the $Q$-branch is too strong. The constraints on $B$ indicate that the carrier of the $\lambda$5797 DIB is a prolate top molecule with 5 to 7 heavy atoms  (for comparison, the rotational constant of HC$_5$N is 1331.3313 MHz \cite{Avery}).

\section{Discussion}

\subsection{Constraints on the size of the carrier of the $\lambda$5797.1~DIB}

\subsubsection{The present analysis: $5~\leq~n~\leq~7$}

 The analysis given above is based on three premises:

(1) The carrier of the $\lambda$5797.1~DIB is a polar molecule as concluded by Oka et al. \cite{2013ApJ...773...42O}

(2) The conspicuous central peak in Figure 1 for $\zeta$~Oph, $\zeta$~Per and $\mu$~Sgr of Kerr et al. \cite{1998ApJ...495..941K} is a $Q$-branch of a parallel transition of a prolate top molecule.

(3) The radiative temperature of the environment is $T$$_r$~=~2.73~K. \\We believe all of them are valid. We further assumed that the carrier of the DIBs is a linear free radical with a $^2\Pi$ ground state and calculated its parallel band spectrum. A similar formalism may work for a parallel transition of a prolate molecule with rotational constants $A$~$\gg$~$B$~$\sim$~$C$ and with $C$$_{2v}$ symmetry such as carbenes, H$_2$C$_n$, or with $A$~$\gg$~$B$ with $C$$_{3v}$ symmetry such as methyl-polyynes, H$_3$C$_n$H. They do not need to be open-shell molecules. In these molecules, the rotational quantum number $K$$_a$ or $K$ plays the role of $\Lambda$ of the $^2\Pi$ linear molecules. Prolate tops with $C$$_s$ symmetry such as CH$_2$CN$^-$ considered by Cordiner and Sarre \cite{Sarre} cannot produce the $Q$-branch since the $b$-type spontaneous emission depopulates the $K$$_a$~=~1 level. Contrary to their argument, rotational levels of ortho and para species are degenerate and they do not keep the $K_a$~=~1 level highly populated.

Although we have only briefly attempted to simulate the $\lambda$5797.1~DIB using the $C$$_{2v}$ and $C$$_{3v}$ prolate tops, we believe that the simulations are similar and that the constraints on $B$ shown in Figure~3 and on $n$ are also applicable for these cases.

 In addition to the constraint from this model calculation, two more constraints can be set from other considerations as discussed below.

 \subsubsection{Upper limit of $n$~$\leq$~6 from the Herschel 36 spectrum}

Oka et al. \cite{2013ApJ...773...42O} explained the anomalous ETRs observable only toward Herschel 36 as due to a high radiative temperature $T$$_r$ and a decrease in rotational constants $B'$~$-$~$B$ upon electronic excitation. This decrease is observed in general for most molecules, but the fractional decrease $\beta$~=~($B'$~$-$$B$)/$B$ is lower for larger molecules as seen in Table 2 of Ref \cite{2013ApJ...773...42O}. Oka et al. set a limit of $\mid$$\beta$$\mid$~$>$~0.015 assuming $T$$_r$$<$~100 K, leading to the constraint $n$~$\leq$~6. This upper limit is not as definitive as that in the previous section in view of the uncertainty of $T$$_r$ toward Herschel 36. For a higher $T$$_r$, the limit $\mid$$\beta$$\mid$~$>$~0.010 may work, which gives $n~\leq~8$ from Table 2 of Ref \cite{2013ApJ...773...42O}. However, such a high $T$$_r$ is not very likely over the long distances through which the DIB molecules are distributed.

\subsubsection{Lower limit of $n$~$\geq$~5 from the stability of molecules against photo-dissociation}

In dense clouds, gas phase reactions make large molecules from small molecules, as evidenced by the relative abundances of series of molecules. For example, the total column densities of cyano-polyynes H-(C$\equiv$C)$_n$-CN for $n$~=~2~$-$~4 toward TMC 1 have been reported to be (in 10$^{13}$~cm$^{-2}$) 5.0, 1.2, 0.32 by Broten et al. \cite{Broten}, 3.3, 1.1, 0.19 by Bell et al. \cite{Bell1}, and 1.1, 1.0, 0.54 by Bell et al. \cite{Bell2}. The values differ due to different treatments of collisions, but we see that the abundance decreases only by a factor of 1~$-$~4 for each additional $-$C$\equiv$C$-$ chain.

The situation is drastically different for diffuse clouds. Oka et al. \cite{OkaC3} reported C$_2$ and C$_3$ column densities toward 14 stars; their ratios are generally larger than 30 and average to $\sim$~45. The high correlation coefficient of 0.93 between them suggests that C$_3$ is produced by chemical reactions
involving C$_2$, but the high C$_2$ to C$_3$ ratio indicates that photo-dissociation and hydrogenation of C$_2$ is much faster than the chemical build-up to C$_3$.
Attempts to detect C$_4$ and C$_5$ failed \cite{OkaC3,MaierC3}. Molecules detected so far in the optical and infrared regions are all diatomics with
the exceptions of H$_3^+$ and C$_3$. Radio observations are more sensitive and have detected triatomics, but H$_2$CO and the ubiquitous $c$-C$_3$H$_2$ and a small amount of $l$-C$_3$H$_2$ are the only polyatomics higher than triatomics \cite{SnowMcCall,Liszt}. All these indicate that abundances of small molecules decrease rapidly with the size of molecules.

On the other hand, the presence of the great many intense DIBs indicates high column densities of large molecules. Using the simple relation between the column density $N$, equivalent width $W_{\lambda}$, and the transition dipole moment $\mu_t$,

\begin{equation}N~=~\frac{3hcW_\lambda}{8\pi^3 \lambda \mu_t^2}, \end{equation}

\noindent the equivalent widths from 1 to 5700 m\AA\ of about 400 DIBs observed toward HD~183143 \cite{2009ApJ...705...32H} and HD~204827 \cite{2008ApJ...680.1256H} indicate high column densities on the order of 5~$\times$10$^{11}$~cm$^{-2}$ to 3~$\times$~10$^{15}$~cm$^{-2}$ for $\mu_t$~=~1~Debye and 5~$\times$10$^{9}$~cm$^{-2}$ to 3~$\times$~10$^{13}$~cm$^{-2}$ for $\mu_t$~=~10~Debye. These are very high column densities and indicate that a large fraction of carbon atoms is locked up in the carriers of DIBs. Moreover, although C$_2$ and C$_3$ are undetectable toward the star HD 183143, while they are most abundant toward the star HD 204827 \cite{OkaC3}, the two sightlines show comparable intensities for $\lambda$5797.1 (see Section 4.2.2.) It is highly unlikely that the carriers of DIBs are produced bottom-up from smaller molecules whose column densities are so low. They must be produced top down from the breakdown of dust by cosmic rays or shocks as proposed by Duley \cite{Duley}, most likely from the mixed aromatic/aliphatic organic nanoparticles (MAONS) proposed by Kwok and Zhang \cite{Kwok}. See also Duley et al. \cite{DuleyZ}. The abundances of the carriers of DIBs must be governed by their stability against photo-dissociation or chemical reactions.

The stability of large carbon molecules C$_n$ has been discussed before in relation to chemical reactions by Freed et al. \cite{Freed} and Clayton et al. \cite{Clayton}. Their stability argument based on the Rice-Ramsperger-Kassel-Marcus (RRKM) statistical model of unimolecular decomposition can also be applied to stability against photo-dissociation. Namely, the energy of an absorbed photon is distributed among the molecule's great many vibrational states. If infrared photons are emitted before the photon energy is concentrated to break a chemical bond, the molecule survives the photo-excitation. As $n$ gets larger, the number of vibrational modes increases rapidly, and the lifetime of C$_n$ grows. The calculations by Freed et al. \cite{Freed}, summarized in their Figure 1, indicate that the lifetime reaches $\sim$~1~s at $n$~=~5. Since infrared spontaneous emission has a typical lifetime of 30 ms, molecules are stabilized by emitting infrared photons. This sets the lower limit of $n$~$\geq$~5.

\subsubsection{$n$~=~7~$-$~8 by Wehres et al.}

W. W. Duley drew our attention to Wehres et al. \cite{Wehres}, in which they assumed a carbon chain and estimated the number of carbon atom to be 7 to 8. This is for an emission band at 6615~\AA\ from the Red Rectangle whose carrier is perhaps the same as that for $\lambda$6613.6 DIB. Their estimate is based on the spacing of three satellite lines which they interpret as due to vibrational excitation in the ground state. This is not directly related to the result of our paper but is worthy of notice.

\subsection{Further considerations for candidate molecule identification}

\subsubsection{Candidate molecules in view of Maier laboratory findings}

John Maier's laboratory has been extremely prolific in studying electronic optical spectra of unstable species. The number of new spectra discovered in his laboratory in Basel is simply staggering. In the history of molecular spectroscopy involving brilliant scientists, no other group with a professor and collaborators has dominated a field so squarely using so many multifaceted experimental techniques.

One issue with the molecular size constraint discussed in the preceding sections is that many 5~$\leq$~$n$~$\leq$~7 molecules as well as larger carbon chain molecules have already been observed in Maier's laboratory and searched for in interstellar space. There are many molecules that have not yet been tested, but one wonders whether such molecules can exist when similar molecular candidates have been rejected. Based on the criteria established by the model, these classes of molecules warrant further investigation as DIB carrier candidates. The following sections discuss the evidence for and against these candidates.

{\textbf{Polar $^2\Pi$ radicals}}

Radicals HC$_n$ first come to mind as candidates based on the criteria established thus far, but the spectra of HC$_5$ and HC$_7$ are perpendicular bands in the wavelength region of interest. They have been observed in the lab and searched for in space \cite{Maier1}. HC$_6$ has been observed in the laboratory and analyzed in detail; it would seem the most promising both because its parallel band appears in the region of interest and because the sign of the spin-orbit splitting constant is negative ($A$~=~$-$~15.04~cm$^{-1}$) \cite{Maier3}. However, a search for it in interstellar space had a negative result \cite{Maier4}.

Radical cations HC$_{n-1}$N$^+$ isoelectronic to HC$_n$ are also candidates. Among these, HC$_5$N$^+$ has a $^2\Pi\leftarrow$$^2\Pi$ band. This molecule has been observed in the laboratory \cite{Maier5} and searched for in interstellar space, also with a negative result \cite{Maier4}. In addition, the cations may not satisfy the stability requirement of Section 4.1.3 because of the rapid dissociative recombination with electrons that is abundant in diffuse clouds.

A series of radicals with a $^2\Pi$ ground electronic state whose electronic transitions have not been observed are the HC$_{n-1}$O molecules observed in the microwave region by the group of Thaddeus and McCarthy \cite{McCarthy}, the group that has dominated microwave studies of carbon chain molecules. The radicals with $n$~$\leq$~4 are bent in the ground state, but the observed microwave spectra for $n$~$\geq$~5 showed indications of being linear molecules with $^2\Pi$ states. The HC$_6$O radical, which has a negative spin-orbit splitting constant $A$, is promising. Theoretical calculations have indicated, however, that the Renner-Teller bent $^2A''$ state is slightly (320 cm$^{-1}$) below the $^2\Pi$ state \cite{McCarthy}. If this is correct, most HC$_6$O would not be in the $^2\Pi$ state because some transition moment must cause the spontaneous emission $^2\Pi$~$\rightarrow$~$^2A''$. For the isovalent series of molecules HC$_{n-1}$S, the ground state is $^2\Pi$ \cite{Endo,Maier6,Maier7}.

Radical anions HC$_{n-1}$N$^-$ isoelectronic to HC$_{n-1}$O radicals are also candidates. Their destruction by C$^+$ may reduce their abundance, although not to as serious an extent as radical cations destroyed by electrons.

There are other radicals with $^2\Pi$ ground states, including ones with Fe, Mg, Si, and S in the chain, but their lower abundances make it less likely that they are carriers of the $\lambda$5797.1~DIB.

{\textbf{C$_{2v}$ and C$_{3v}$ molecules}}

Cumulenes, linear H$_2$C$_n$, have been observed in the microwave region by Thaddeus's group in the laboratory for $n$~=~3~$-$~9 \cite{Thaddeus1,Thaddeus2,Thaddeus3,Thaddeus4} and in interstellar space for $n$~=~3 and 4 \cite{Thaddeus5,Thaddeus6,1991PASJ...43..607K} and 6 \cite{Thaddeus7}. They all have large permanent dipole moments; 4.1, 4.6, 5.9, 6.2 Debye for $n$~=~3, 4, 5, 6, respectively \cite{Defrees,Dykstra,McLean}. H$_2$C$_n$ with even $n$ is much more abundant than with odd $n$. The abundance of H$_2$C$_6$, which is most relevant to discussions in this paper, is reported to be less than that of H$_2$C$_4$ by a factor of 25 in TMC-1, but this ratio may well be reversed in diffuse clouds due to the stability against photodissociation discussed in Section 4.1.3. The longest cumulene observed in the optical region is H$_2$C$_3$, which was proposed to be the carrier of the $\lambda$5450~DIB by Maier et al. \cite{Maier8}. Cumulenes with higher $n$ may well be viable candidates.
A recent theoretical paper, however, predicts low oscillator strengths \cite{Zhang} which may make cumulenes unlikely candidates (see Section 4.2.2 below).

The incredibly prolific lab of Thaddeus and McCarthy has produced cumulene derivatives such as H$_2$C$_4$N, H$_2$C$_6$N, H$_2$C$_5$H, H$_2$C$_7$H and many C$_{3v}$ molecules \cite{Thaddeus8}, all of which could be candidates for the carrier of the $\lambda$5797.1 DIB if they have electronic transitions in the right wavelength region.

\subsubsection{Required high column density}

When Maier et al. proposed H$_2$C$_3$ as the carrier of the $\lambda\lambda$4881 and 5450 DIBs, the main objection was that the required H$_2$C$_3$ column densities on the order of 5~$\times$~10$^{14}$ toward HD 183143 and 2~$\times$~10$^{14}$ toward HD 206267 were far too high \cite{Oka,Krelowski,Araki,Liszt}. Araki et al. \cite{Araki} searched for the 5$_{1,5}$~$\rightarrow$~4$_{1,4}$ rotational emission at 102.99238 GHz toward HD 183143. Based on the non-detection of the line, they set the upper limit of the H$_2$C$_3$ column density to be 1.1~$\times$~10$^{13}$~cm$^{-2}$ or 2.0~$\times$~10$^{13}$~cm$^{-2}$ for the excitation temperatures of 10 K and 60 K, respectively, assumed by Maier et al. \cite{Maier8}. Their value, however, does not set a valid upper limit. The frequency of 103 GHz ($\sim$~4.9 K) is far too high to be radiatively pumped at $T_r$~=~2.73 K. In addition, the spontaneous emission rate, assuming a permanent dipole moment of 4.1 Debye \cite{Defrees}, would be 9.3~$\times$~10$^{-5}$~s$^{-1}$, which is 1000 times higher than the average collision rate of $C$~=~10$^{-7}$~s$^{-1}$. The 5$_{1,5}$ level won't be populated in diffuse clouds even if the column density of H$_2$C$_3$ is high.

Liszt et al. \cite{Liszt} observed the lowest $R$-branch absorption line 1$_{01}$~$\leftarrow$~0$_{00}$ of para-H$_2$C$_3$ at 20792.59 MHz and determined the column density to be on the order of 10$^{11}$~cm$^{-2}$, three orders of magnitude smaller than that claimed by Maier et al. Although the sightlines observed are not the same as used by Maier et al., approximate proportionality between molecular abundance and the reddening $E_{B-V}$ is used to normalize the column density. The 0$_{00}$ level is certainly populated and this gives the correct column density. A surprising observation in their paper is that the fractional abundances of molecules in Table 4 are similar for diffuse clouds and for TMC-1. If this similarity holds for larger molecules, it will contradict the speculation in Section 5.1.3 of this paper that DIB molecules are produced not by chemical reaction in space but from the breakdown of MAONS. \emph{In order to check this it is important to observe large molecules like H$_2$C$_6$ in absorption in the centimeter region.}

In any case, the large column densities required to account for the observed large equivalent widths will present a problem whatever the carrier molecule is. The equivalent width of the $\lambda$5797.1 DIB has been reported to be 199.0 m\AA\ toward HD 204827 \cite{2008ApJ...680.1256H} and 186.4 m\AA\ toward HD 183143 \cite{2009ApJ...705...32H}. From Equation 7, these values correspond to column densities of 8.3~$\times$~10$^{13}$~cm$^{-2}$ and 7.7~$\times$~10$^{13}$~cm$^{-2}$, respectively, if $\mu_t$ is 1 Debye. The column densities are lower by a factor of 100 if $\mu_t$ is 10 Debye. The intensity of an electronic transition is often given in terms of the oscillator strength

\begin{equation}f~=~\frac{8\pi^2 m_e}{3he^2} \nu\mu_t^2~=~4.701~\times~10^{-7}~\tilde{\nu}\tilde{\mu_t}^2~=~8.109~\times~10^{-3}\tilde{\mu_t}^2, \end{equation}

\noindent where $\tilde{\nu}$ and $\tilde{\mu_t}$ are the frequency of the DIB in cm$^{-1}$ and the transition dipole moment in Debye, respectively, with the last formula applying to the $\lambda$5797.1 DIB. This indicates that $f$~=~1 corresponds approximately to $\tilde{\mu_t}$~=~10 Debye. Thus if $f$~$\sim$~1 (e.g. Table 2 of Ref \cite{Maier9}), the required column density is on the order of 10$^{12}$~cm$^{-2}$, which may be achievable in diffuse clouds.

\bigskip

This paper is dedicated to John P. Maier in appreciation of the inspiration he has given spectroscopists and molecular astrophysicists through his great work over many years.
We thank Daniel E. Welty for providing reduced spectra of 20 Aql and Herschel 36 based on data obtained from the ESO Science Archive Facility. We are grateful to M. Araki, W. W. Duley, L. M. Hobbs, M. C. McCarthy, J. P. Maier, P. J. Sarre, P. Sonnentrucker, G. A. H. Walker, D. E. Welty, A. Witt, and D. G. York for reading our draft and giving helpful comments. We also thank two anonymous referees for comments which improved the presentation of this paper. This research was supported by NSF grant 1109014.

\bibliographystyle{tMPH}
\bibliography{dibs}

\appendix
\section{Re-examination of the Anomalous Her 36 DIBs}

Our use of the $^2\Pi \leftarrow$~$^2\Pi$ parallel band for the $\lambda$5797.1 DIB in this paper contradicts the use of the $^2\Pi \leftarrow$~$^2\Sigma$ perpendicular band by Oka et al. \cite{2013ApJ...773...42O} to account for the ETR toward the star Herschel 36. The perpendicular band was used since a parallel band cannot simulate the ETR if a uniform high radiative temperature $T_r$ is assumed. An example of a simulation of the $\lambda$5797.1 DIB as a parallel band using $T_r$ = 25 K, $\lambda_0=5797.25$ $\angstrom$, and the remaining
parameters fixed at the values in Model A of this paper is presented in Fig. A1. The two distinct peaks differ from the observed single
peak and this dilemma cannot be fixed as long as we use a single $T_r$.

In actuality, $T_r$ varies along the line of sight from very high $T_r$ near Her 36 SE to $T_r$~=~2.73~K at greater distances from this infrared source. The value of $T_r$ used by Oka et al. \cite{2013ApJ...773...42O} is an effective temperature after a very complicated average. Since taking into account the variation of $T_r$ along the line of sight is a difficult task, here we add spectra calculated with $T_r$~=~30~K and $T_r=2.73$~K. Fig. A2 provides an example. The ETR is reproduced by the band at the higher temperature, and the double-peaked structure of the parallel band at the higher temperature
is ``filled in'' by the transition occurring at a lower temperature. Thus the ETR on $\lambda$5797.1~DIB can be reproduced by a perpendicular band as well as by a parallel band as long as the fraction of the variation of the $B$ rotational constant $\beta$ is sufficiently high.

\begin{figure*}
\begin{center}
\includegraphics[scale=0.6]{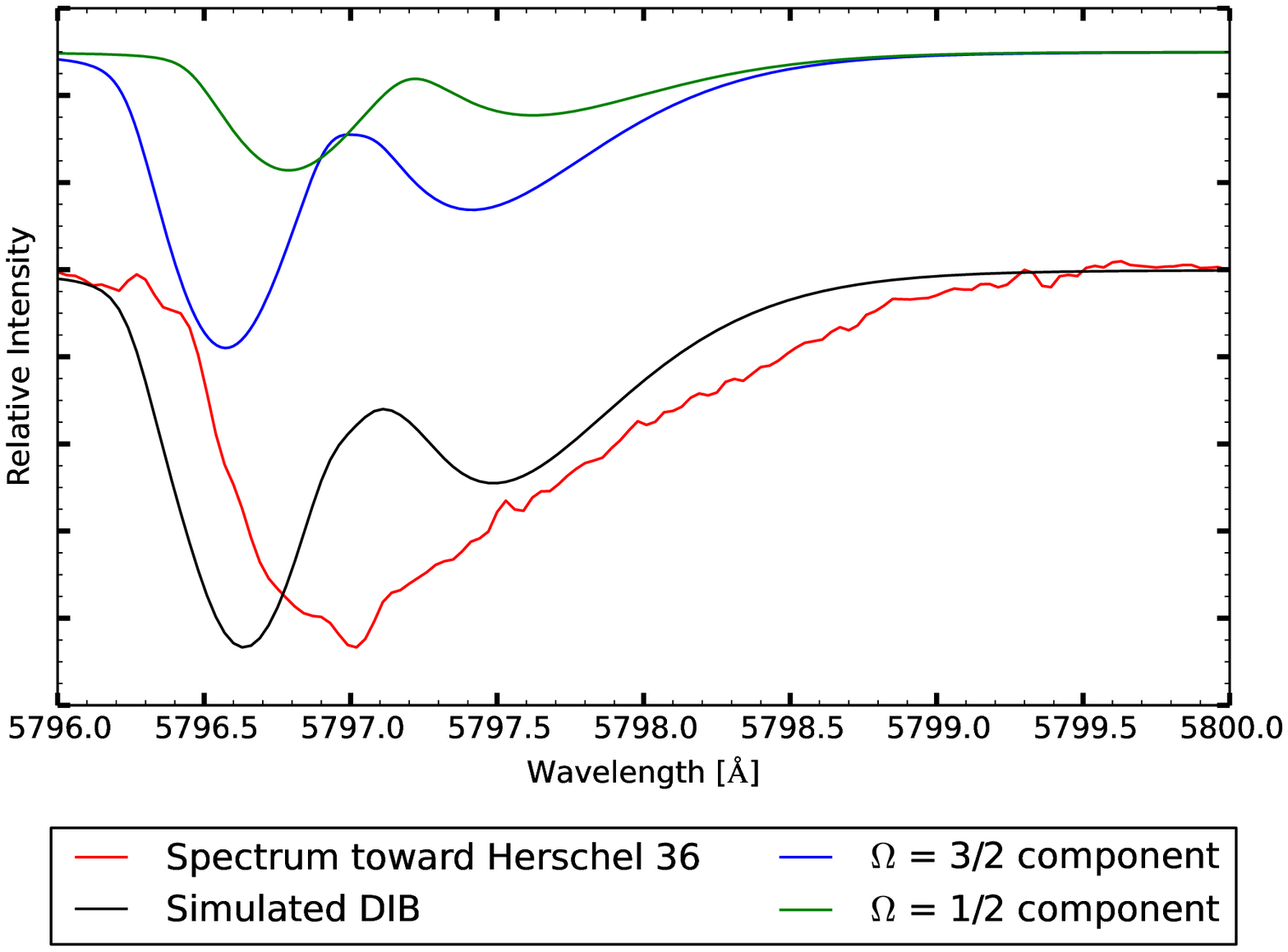}
\end{center}
\label{fig4}
\caption{Simulation of the anomalous $\lambda$5797.1 DIB toward Her 36 using a $^2 \Pi \leftarrow$
$^2\Pi$ parallel transition, $T_r$ = 25 K, $\lambda_0=5797.25$ $\angstrom$. All other model input values equal
to those of Model A. The parallel band based on a single $T_r$ does not reproduce the observed spectrum. The $R=48000$ Her 36 spectrum (courtesy of Dan Welty) was retrieved from the ESO Science Archive Facility.}
\end{figure*}

\begin{figure*}
\begin{center}
\includegraphics[scale=0.6]{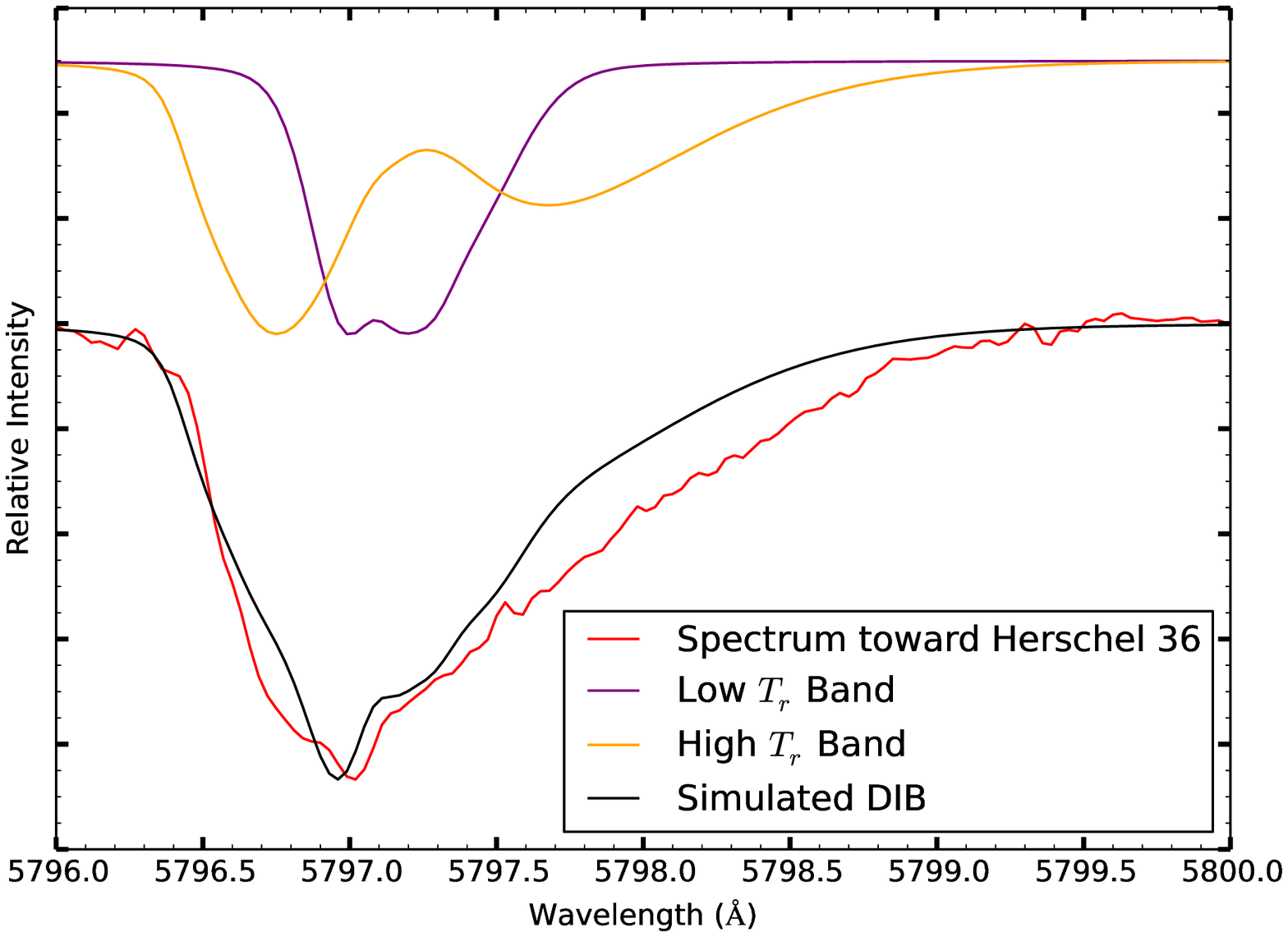}
\end{center}
\label{fig5}
\caption{Alternative simulation of the anomalous $\lambda$5797.1 DIB toward
Her 36 using the parallel $^2 \Pi \leftarrow$ $^2\Pi$ transition using two radiative temperatures $T_r$ = 2.73 K for narrow component and  $T_r$ = 30 K for the broader component. See appendix for further explanation.}
\end{figure*}

\end{document}